# Extending a Microsimulation of the Port of Dover

*Chris Roadknight and Uwe Aickelin*
Intelligent Modelling & Analysis Research Group (IMA)
School of Computer Science
The University of Nottingham
Jubilee Campus
Wollaton Road
Nottingham NG8 1BB
{cmr,uxa}@cs.nott.ac.uk

## ABSTRACT

Modelling and simulating the traffic of heavily used but secure environments such as seaports and airports is of increasing importance. This paper discusses issues and problems that may arise when extending an existing microsimulation strategy. This paper discusses the simulations and how extensions of these simulations can aid planners with optimal physical and operational feedback. Conclusions are drawn about how microsimulations can be moved forward as a robust planning tool for the 21st century.

## 1    INTRODUCTION

Traffic microsimulation software is becoming increasingly complex, parameterized and configurable. Regardless of how graphically realistic the end product may appear, the core statistics generated by the simulation still needs to be validated and verified. Events and statistics that show up when simulations are tested must also appear in real life and vice versa. Real world validation of simulation results can be an expensive, time consuming, subjective and erroneous process and deciding exactly how much validation to commission is usually an imprecise art. Weighing up the cost/reward ratio of validation is an important but non-trivial process. Any changes to the layout of the port will affect throughput and resulting impact on traffic flow on the approaches to Dover. These changes might be in the form of increasing or decreasing the number of lanes, changing the document checking protocols, increases in security checks or expansion of the site.

Traffic Microsimulations use a discrete event [5] approach to the movement of vehicles over time where the behavior of a system is represented as a chronological sequence of events. Each event occurs at a unique instant in time, with each new instance of the system viewed as new state. They combine this with some degree of agent based behavior where elements within the simulation have a set of parameters and policies that they use to come to decisions. If validation is not properly performed, a traffic simulation model may not provide accurate results and shouldn't be used to make important decisions with financial, environmental and social impacts.

The research in this paper involves simulations and real worth data from the the Port of Dover. It was chosen for this research as it is the most important trading route between the UK and mainland Europe, has an intricate and multilevel layout and has a substantial amount of existing data and simulations. Over the past 20 years, the number of road haulage vehicles (RHVs) using the Port of Dover has more than doubled to over 2.3 million [4]. Looking ahead over the next 30 years, both the Port and UK Government have forecast substantial growth in RHV freight traffic. Approximately 3 million tourist vehicles also pass through the ferry port annually making it a key European and global tourist gateway.

This paper sets out to identify the performance and characteristics of a microsimulation approach to closed system vehicle simulation with particular reference to the stability and reproducibility of the simulations when they are modifed. The next section of this paper outlines existing data, statistics and graphs for the Port of Dover, Section 3 discusses the simulation software package VISSIM, Section 4 introduces a novel validation procedure which is tested at the Port of Dover, Section 5 discusses an extension of the validated simulation and Section 6 offers some conclusions. We attempt to answer questions about the usefulness of microsimulations with relation to the variability and accuracy of the simulations when compared to real world data.

## 2      SIMULATING THE PORT OF DOVER PAPER

Microsimulations of the Port of Dover exist [1,3] and have been used better understand the flow around the Port as well as the impact of increased future load. It has been shown that care must be taken when designing these simulations to ensure the correct balance of agent intelligence, solution transparency and statistical representation. Discrete event based simulation environments such as VISSIM provides agents with detailed driver behavior [2] where driver parameters can be selected from a known distribution. VISSIM [2] is a leading microscopic simulation program for multi-modal traffic flow modeling. It has a high level of vehicle behaviour detail that can be used to simulate urban and highway traffic, including pedestrians, cyclists and motorised vehicles. It is a highly parameterised design system that allows a lot of flexibility. VISSIM models provide detailed estimates of evolving network conditions by modeling time-varying demand patterns and individual drivers' detailed behavioral decisions. Several model inputs (such as origin flows) and parameters (car-following and lane-changing coefficients) must be specified before these simulation tools can be applied, and their values must be determined so that the simulation output accurately replicates the reality reflected in traffic measurements..

## 3      EXTENDING THE SIMULATION

Y An existing simulations of the Port of Dover exist and have been evaluated and discussed at length [3]. It is a closed system where the outward in inward bound traffic does not mix. Here we will only discuss the outbound vehicles that are en route to France. Base line existing metrics such as trip time and queue length will be used as a starting point and comparison. The issue of automated ticketing of drivers has been investigated with a view to deciding if the option of adding additional lanes for some drivers with tickets enabled for automated ticketing were suitable designs. A few assumptions are made at the start:

1. Automated ticketing has the same processing time distribution as the manual ticketing, wait time was normally distributed with a mean wait time of 77 seconds with a standard deviation of 50 seconds. Discussion with the Port of Dover suggests that automated ticketing should generally have a lower average processing time, though possibly with a longer tail.

2. Both options use a single filter lane approach to a five lane plaza, the 2 options have different lane lengths

3. The decision point for choosing lanes is at the post-weighbridge merge point where HGVs rejoin the tourist flow. VISSIMs standard vehicle lane selection processes operates here to ensure vehicles enter safely.

4. The flow used was a peak flow (~6 vehicles/minute) accurately represented in 2 minutes segments for 90 minutes. The observed HGV/ Car ratio was used.

5. Average trip time was used as the assessment metric. It must be remembered that this includes vehicles that both took the automated ticketing route and those that didn't.

Firstly we simulated what would happen if a fraction of ALL vehicles made the decision to take the automated ticketing route. For this we used the longer Option 1 solution, this can accommodate up to 10 cars per lane (50 total) but fewer HGVs (figure 1). Figure 2 shows the average trip times at different times for values of between 10 and 30% against the original baseline (ie. 0%). It can be seen that in terms of average trip time there is a clear advantage of between 85-95 seconds in assigning 10-20% of vehicles to Automated ticketing but a significant disadvantage when 30% of vehicles chose the Automated ticketing route. The time gain is due to the fact that the new route is shorter and also that 5 new (automated) kiosks are opened therefore putting less pressure on the existing kiosks

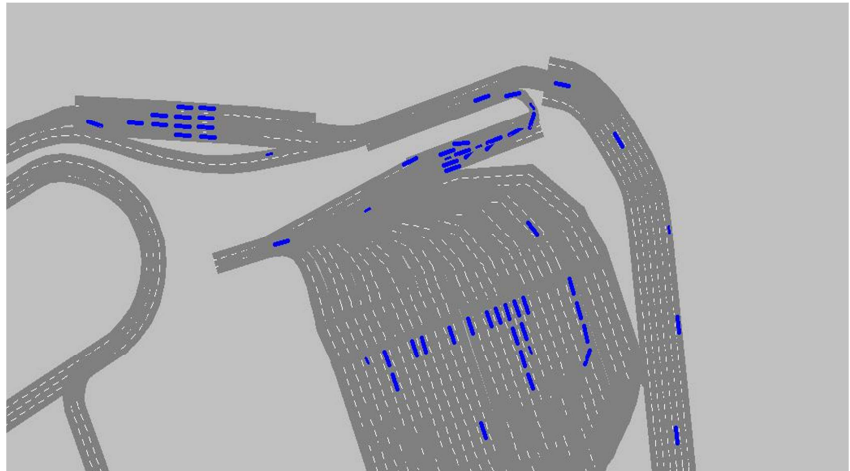

**Figure 1**. *Addition of an automated ticketing lane for all vehicles (Option 1)*

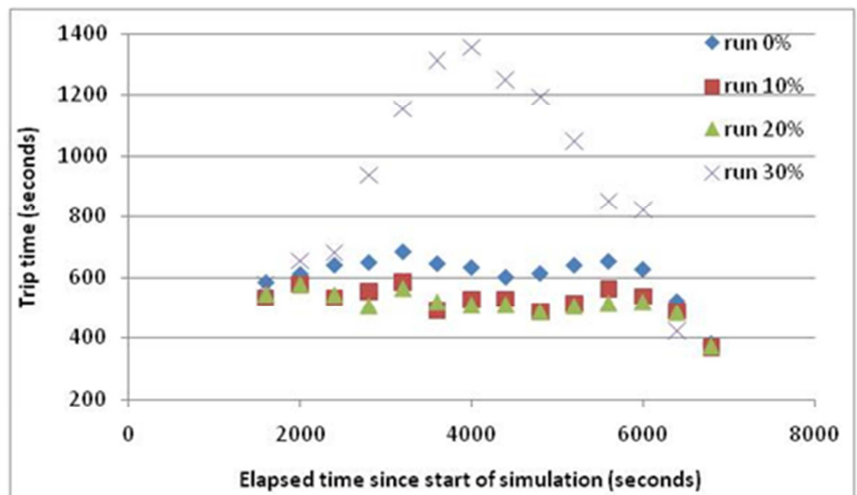

**Figure 2.** *Average trip times (in 400 second bins) for different fractions of Automated ticketing*

The added trip time at 30% is due to congestion at the Automated ticketing lane delaying both sets of vehicles. There is a significant amount of variation between runs with different random number seeds, especially at upper and lower automated ticketing values. (Fig 3). This suggests that at these levels the system is operating at a delicate point where small changes in flow variability have a large impact on trip time.

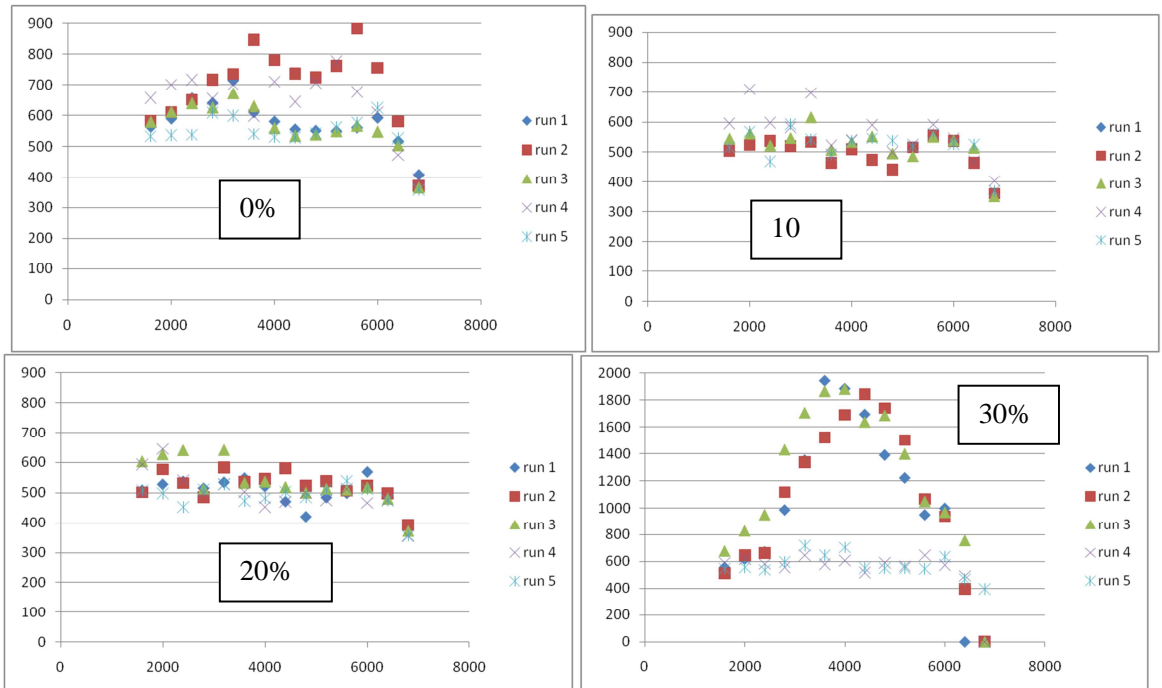

**Figure 3.** *Variation between runs for different levels of automated ticketing adoption*

If we now assume that the only vehicles that can use the automated ticketing lanes are **cars** then the number of vehicles that can pass through the automated ticketing while providing an overall reduction in trip time reaches 100%, this is also true if shorter lanes are used (maximum of 5 cars per lane, 25 in total) both options provided significantly lower trip times than the existing non-automated ticketing approach). Figure 4 shows this as "50 5" and "25 5". 50 and 25 represents the number of vehicles that can be accommodated in the new automated ticketing queues the 2 options and 5 represents the number of ticketing kiosks that are open. Only when the number of automated ticketing kiosks is reduced to 2 and 3 do delays increase ("25 2" means there is space for 25 cars and 2 kiosks are open and "25 3" means there is space for 25 cars and 3 kiosks are open) there is also more variability between runs.

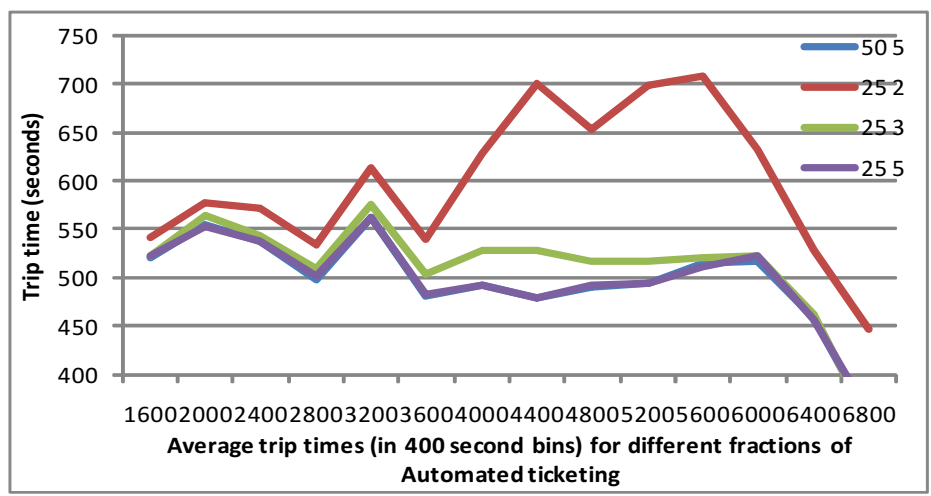

**Figure 4.** *Comparison of trip times for a fully open Option 1 (50 5) and a fully and partial opened Option 2 (25 5, 25 3, 25 2)*

## 4    CONCLUSIONS

When building a microsimulation great care must be taken to ensure each component is as accurate as possible as small errors in design can lead to disproportionately large errors. This is especially the case if actual behavior is replaced with probabilistic approaches, while these can ensure representative statistics they can also introduce gross errors when coupled with strict lane discipline and can also be an example of overcalibration. There is a requirement in an agent based simulation to have appropriately intelligent agents that best reflect actual behavior without introducing significant overheads in terms of complexity and hardware requirements. Having agents with representative behavior reduces the need to overcalibrate the system by using popular methods such as probabilistic routing.

Simulations can be constructed to such an accuracy as to completely mimic the situation used as an example, but when this simulation requires extension or modification there may be situations of where the extension is not accurate due to overcalibration of the initial simulation. It is easier to create overcalibration errors using modern, componantised simulation software where each individual element can be highly configured to be representative of the isolated sub-system without requiring any system wide validity.

The usefulness of a microsimulation of somewhere like Dover is demonstrated by adding an extension, in this case automated ticketing and testing behavior under different circumstances. The results of this are difficult to validate due to the predictive nature of the design whereby the actual scenario hasn't been introduced yet but future work will involve returning to the site once automated ticketing is used to evaluate accuracy.

## ACKNOWLEDGMENTS

An acknowledgements section may be added, if required, to acknowledge the contribution of other research work. The acknowledgements section should be placed between the main text of the paper and before the References section (or any appendices included).

## REFERENCES

[1] C Roadknight and U Aickelin. "Validation of Traffic Flow Simulations through the Port of Dover" Proceedings of ICMSC 2010, p357-361.
[2] M. Fellendorf and P. Vortisch. "Validation of the microscopic traffic flow model VISSIM in different real-world situations". 80th Meeting of the Transportation Research Board. Washington, D.C., January, 2001.
[3] C Roadknight, U Aickelin and G Sherman. "Validation of a Microsimulation of the Port of Dover", Journal of Computational Science, ISSN 1877-7503, 10.1016/j.jocs.2011.07.005
[4] http://www.doverport.co.uk/?page=PortDevelopment
[5] S. Robinson (2004). Simulation - The practice of model development and use. Wiley.

## AUTHOR BIOGRAPHIES

**CHRIS ROADKNIGHT** After completing my PhD on Neural Networks for Biological Data Modelling, Chris joined the then Multimedia Research Group at BT working on modelling the performance of WWW caches.  Since then he has specialised in the areas of distributed computing, biologically inspired AI, their applications and system modelling. Major research areas include Pervasive ICT, A-Life, lightweight artificial intelligence embodiment and wireless sensor networks. Later roles involved technical design and development of prototype AI systems for sensor networks. He has written or co-authored over 30 papers and 3 patents, been involved with W3C steering groups and on the review panel for several conferences. Review articles on this work have also appeared in more mainstream publications such as New Scientist and BBC online. After leaving BT he pursued a research career initially at Lancaster University and now at Nottingham University, working as a Research Fellow in the areas of simulation and AI. http://ima.ac.uk/roadknight

**UWE AICKELIN**

Uwe Aickelin is EPSRC Advanced Research Fellow and Professor of Computer Science at The University of Nottingham, where he leads one of its four research groups: Intelligent Modelling & Analysis (IMA). His long-term research vision is to create an integrated framework of problem understanding, modelling and analysis techniques, based on an inter-disciplinary perspective of their closely-coupled nature. A summary of his current research interests is Modelling, Artificial Intelligence and Complexity Science for Data Analysis.closely-coupled nature. A summary of his current research interests is Modelling, Artificial Intelligence and Complexity Science for Data Analysis http://ima.ac.uk/aickelin/